\documentclass[%
 aip,
 amsmath,amssymb,
 reprint,%
]{revtex4-1}

\usepackage{graphicx}
\usepackage{dcolumn}
\usepackage{bm}
\usepackage{bm}
\usepackage[utf8]{inputenc}
\usepackage[T1]{fontenc}
\usepackage{mathptmx}
\usepackage{algorithm}
\usepackage{algpseudocode} 
\usepackage{etoolbox}
\usepackage{xcolor}
\usepackage{amsmath}
\usepackage{placeins}
\usepackage{amssymb}
\usepackage{subfig}
\usepackage[]{graphicx}

\makeatletter
\def\@email#1#2{%
 \endgroup
 \patchcmd{\titleblock@produce}
  {\frontmatter@RRAPformat}
  {\frontmatter@RRAPformat{\produce@RRAP{*#1\href{mailto:#2}{#2}}}\frontmatter@RRAPformat}
  {}{}
}
\begin{document}

\preprint{AIP/123-QED}

\title[Smooth Particle Mesh Ewald-integrated stochastic Lanczos Many-body Dispersion algorithm]{Smooth Particle Mesh Ewald-integrated stochastic Lanczos Many-body Dispersion algorithm}

\author{Pier Paolo Poier$^*$}
\email{pier.poier@sorbonne-universite.fr}
\affiliation{Sorbonne Université, Laboratoire de Chimie Théorique, UMR 7616 CNRS, 75005, Paris, France}
\author{Louis Lagardère}
\affiliation{Sorbonne Université, Laboratoire de Chimie Théorique, UMR 7616 CNRS, 75005, Paris, France}
\affiliation{Sorbonne Universit\'e, IP2CT, FR 2622 CNRS, Paris, France}
\author{Jean-Philip Piquemal$^*$}
\affiliation{Sorbonne Université, Laboratoire de Chimie Théorique, UMR 7616 CNRS, 75005, Paris, France}
\affiliation{The University of Texas at Austin, Department of Biomedical Engineering, TX, USA}
\email{jean-philip.piquemal@sorbonne-universite.fr}

\date{\today}

\begin{abstract}
We derive and implement an alternative formulation of the Stochastic Lanczos algorithm to be employed in connection with the Many-Body Dispersion model (MBD). Indeed, this formulation, which is only possible due to the Stochastic Lanczos' reliance on matrix-vector products, introduces generalized dipoles and fields. These key quantities allow for a state-of-the-art treatment of periodic boundary conditions via the $\mathcal{O}(N\log(N))$ Smooth Particle Mesh Ewald (SPME) approach which uses efficient fast Fourier transforms. This SPME-Lanczos algorithm drastically outperforms the standard replica method which is affected by a slow and conditionally convergence rate that limits an efficient and reliable inclusion of long-range periodic boundary conditions interactions in many-body dispersion modelling. The proposed algorithm inherits the embarrassingly parallelism of the original Stochastic Lanczos scheme, thus opening up for a fully converged and efficient periodic boundary conditions treatment of MBD approaches. 
\end{abstract}
\maketitle
\section{\label{sec:intro}Introduction}
Electron correlation is one of the most fascinating and difficult phenomenon to model. Dispersion in particular originates from the long-range electronic correlation among distant electron densities and represents the purely attractive contribution in van der Waals interactions. These are ubiquitous in nature: they can be for example observed in milk as they drive the formation of lipid droplets that, through light scattering, give to milk its typical white color. Geckos and spiders, on the other hand, also take advantage of dispersion for supporting their entire weight on smooth vertical surfaces.\\
From the microscopic point of view, dispersion interactions are crucial in many processes driven by non-covalent phenomena such as protein folding, protein-protein interactions, supra molecular and inter-molecular interactions in general.\\
An exact modelization of dispersion requires the
analytical solution of the electronic Schrödinger equation, which is unfortunately impossible for practical cases. In the past decades, very accurate numerical wave function-based quantum chemical methods have been developed to tackle electron correlation, thus implicitly capable of describing dispersion and intermolecular interactions.\cite{olsen_bible,frank} These methodologies, however, can only be applied to molecules composed of very few atoms, thus preventing the study of chemically and biologically relevant systems.\\
The advent of Density Functional Theory (DFT) represents a milestone in quantum chemistry as it provides a cheap way of including electronic correlation, as its computational cost is similar to that of the Hartree-Fock method. Nevertheless, the intrinsic local nature of common exchange-correlation functionals, makes DFT inadequate for describing long-range correlation effects, thus dispersion.  
To retain the DFT scaling benefits, extensive efforts have been spent in the past years in developing dispersion corrections able to improve the DFT capability of describing intermolecular interactions, crucial in material design and molecular modelling in general.\\
Many of these correction techniques rely on simple empirical pairwise treatments of dispersion, similar to those embraced in force fields. Their simplicity, together with the negligible computational cost and the good accuracy improvement, made possible for these methods to be included in most of the quantum chemistry softwares.\cite{grimmechemrev}\\ 
Despite their large diffusion, these pairwise corrections completely neglect the many-body nature of dispersion interactions inherited from the long-range electronic correlation on the basis of these phenomena.\\
In recent years, the interest towards  Many-Body Dispersion correction models has risen\cite{donchev}. In particular the MBD@rSCS model by Tkatchenko, Di Stasio and Ambrosetti, together with its variations, has become especially popular by virtue of its high accuracy obtained despite of the absence of empirical parameters except for a single range-separation parameter for the coupling between the long-range MBD energy and the chosen DFT functional.\cite{mbd_rsscs,fracpol,mbdsapt}\\
The MBD@rsSCS model can be summarized as follows. First, a set of atomic dipole polarizabilities are obtained from the partitioning of the molecular electron density or, alternatively, retrieved from a deep-neuronal network as recently proposed.\cite{dnnmbd,mbdq} Secondly, the polarizabilies are made frequency-dependent via Padé approximation and subsequently a Dyson-like self-consistent screening linear equation is solved for a selected set of frequencies. Lastly, the set of screened frequency-dependent polarizabilities are used as key quantities in building the MBD interaction matrix which spectrum is used to express the final many-body dispersion energy.\\
Compared to the $\mathcal{O}(N^4)$ scaling of Kohn-Sham equations' resolution, the MBD@rsSCS model involves a small additional computational cost. However, for increasingly large systems, the $\mathcal{O}(N^3)$ scaling of the diagonalization procedure becomes no longer negligible and, it can even become a burden if coupled to $\mathcal{O}(N)$ DFT methods.\\
Recently, we have proposed and implemented an alternative resolution of the MBD key equations that overcomes this scaling issue that is based on the state-of-art Stochastic Lanczos (SL) trace estimation.\cite{stoclancz} Due to the the sparsity of the matrices involved, it exhibits linear-scaling with the system size. The proposed stochastic Lanczos  MBD approach (SL-MBD) further benefits from an embarrassingly parallel implementation arising from its stochastic nature and this allows for reaching system sizes of hundred thousands atoms within a few minutes' time. \cite{mbd_stoch} \\
Compared to a simple pairwise description, this many-body treatment of dispersion interactions in systems such as solvated proteins has revealed a higher degree of delocalization as well as a collective solute-solvent character leading to remarkable long-range interactions.\cite{plasmonicmbd} The potentially longer-range of MBD interactions stresses the importance of the inclusion of a coherent full periodic boundary condition (PBC) treatment, especially in highly ordered and periodic systems. In this direction, recent efforts have been spent in past years. Bucko and co-workers have provided a method expanding over the Brillouin cell that introduced consistent improvements compared to the standard replica method used to include long-range periodic boundary conditions effects.\cite{buco_pbc}\\
By virtue of the above mentioned long-range nature of MBD interactions, it is of broad interest to generalize the SL-MBD approach to a full PBC treatment. However, the quadratic-scaling approaches typically employed in connection to MBD models are clearly not suitable to be integrated in the the SL-MBD methodology for both memory requirements and computational efficiency due to the large systems targeted. A more sophisticated approach has therefore to be developed.\\
In the context of long-range electrostatics modelling, this scaling limitation was addressed via Ewald summation techniques, as they formally scale as $\mathcal{O}(N^2)$ but a proper optimization lowers the factor to $\mathcal{O}(N^{3/2})$. Ewald summation techniques replace the original conditionally convergent energy summation with a direct and reciprocal space absolutely convergent ones consisting of a real and reciprocal summations as well as a self interaction term. The Particle Mesh Ewald (PME) method proposed by Darden, York and Pedersen, drastically improved Ewald summation technique's associated performance.\cite{darden} Its idea relies on the efficient calculation of the reciprocal space energy contribution thanks to fast Fourier transforms scaling as $\mathcal{O}(N\log(N))$. The PME method with its different variants (especially the Smooth Particle Mesh Ewald (SPME)\cite{spme}, has become the standard algorithm implemented in nearly all the most efficient Molecular Dynamics packages thanks to its scaling features although alternative but related methods also exist.\cite{york} \\
 In this work, we derive and present a modification of the SL-MBD method based on a PME treatment of periodic boundary conditions. The resulting Smooth Particle Mesh Ewald stochastic Lanczocz (SPME-SL) MBD approach is suitable for large systems as it exhibits the typical $\mathcal{O}(N\log(N))$ scaling inherited from the PME method.\\
In the next section, we review the MBD model as well as the stochastic Lanczos method in its standard form. A theory section is then dedicated to the derivation of the modified SPME-based Lanczos quadrature scheme followed by a section dedicated to numerical results where the computational performances of the method are discussed and compared to the ones of the standard replica method.
\section{\label{sec:reviewmbd}Review of the MBD and SL-MBD}
The MBD model is based on the idea that a molecule is described as a set of interacting quantum harmonic oscillators, which Hamiltonian is shown in eq.\eqref{ham}, $\mathbf{d}_i=\sqrt{m_i}\boldsymbol{\xi}_i$ being the mass-weighted dipole moment displaced by the vector $\boldsymbol{\xi}_i$ from its equilibrium position. $\alpha_i(0)$ and $\omega_i$ represent the model's key parameters and correspond to the static dipole polarizability and characteristic excitation frequency respectively. 
\begin{equation}\label{ham}
\begin{split}
\hat{H}_\text{MBD}&=\frac{1}{2}\sum_{i={1}}^{N}(-\hat{\nabla}^2_{\mathbf{d}_i}+\mathbf{d}_i^\dagger \mathbf{V}_{ii}\mathbf{d}_i)+\sum_{i>j}\mathbf{d}_i^\dagger \mathbf{V}_{ij} \mathbf{d}_j\\
\mathbf{V}_{ij}&=\mathbf{I}_3\delta_{ij}{\omega}_i^2+(1-\delta_{ij}){\omega}_i{\omega}_j\sqrt{{\alpha}_i(0){\alpha}_j(0)} \mathbf{T'}_{ij}(\beta)
\end{split}
\end{equation}
These parameters are obtained from \emph{ab initio} data as the atom-in-molecule (AIM) polarizability is typically retrieved via partitioning of the electron density while $\omega_i$ is defined in terms of accurate free atoms quantities.\cite{politzer,johnson_vol,tksche}\\
We note that these AIM polarizability parameters can be screened by solving a Dyson-like equation\cite{mbd_rsscs} that can be solved extremely efficiently\cite{mbd_stoch}, however, we will not discuss this 
 in the present work as the presented algorithm is general and does not depend neither on the choice of AIM polarizabilities nor on their screening. We further mention that recently Johnson and coworkers have analyzed the sensitivity of the screening procedure for selected systems.\cite{sensitivity}\\
The $\mathbf{T'}_{ij}(\beta)$ term is built from the pure point dipole-dipole interaction tensor for the $ij$ atom pair that is screened via a damping function $s(R_{ij};\beta)$ depending on the interactomic distance $R_{ij}$ and the single range-separation parameter $\beta$ typically optimized for the correspondent DFT functional to be dispersion-corrected, $\mathbf{T'}_{ij}(\beta)=s(R_{ij};\beta)\mathbf{T}_{ij}$. Recently the MBD model was generalized to higher than dipole interactions\cite{mbdq,ambrogio}, however, here we will only consider the dipole-dipole interaction case. For the explicit expression of $\mathbf{T'}$ we refer to the work in reference.\cite{mbd_stoch} The eigenvalues ($\lambda_i$) of the MBD interaction matrix $\mathbf{V}$, shown for the $ij$ block in eq.\eqref{ham}, are required to obtain the MBD energy $\mathcal{E}_\text{MBD}$ via the plasmonic formula shown in eq.\eqref{plasmonic} that represents the correlation energy of the interacting fluctuating dipoles.
\begin{equation}
    \label{plasmonic}
    \mathcal{E}_\text{MBD}=\frac{1}{2}\sum_{i=1}^{3N}\sqrt{\lambda_i}-\frac{3}{2}\sum_{i=1}^N\omega_i
\end{equation} 
The solution of eq.\eqref{plasmonic} is bound to the $\mathcal{O}(N^3)$ scaling of the diagonalization step that, as mentioned earlier, strongly limits the applicability of the method to large systems.\\
The SL-MBD method bypasses the diagonalization of $\mathbf{V}$ by exploiting the alternative but equivalent expression of the plasmonic formula, eq.\eqref{plasmonic2}, where the sum over the whole spectrum of $\mathbf{V}$ is rewritten in term of its trace, that is invariant under any change of basis, namely $\sum_{i=1}^{3N}\sqrt{\lambda_i}=\text{Tr}{(\sqrt{\boldsymbol{\Lambda}})}=\text{Tr}{(\sqrt{\mathbf{V}})}$ where $\boldsymbol{\Lambda}$ is the diagonal form of $\mathbf{V}$ obtained via the unitary transformation $\boldsymbol{\Lambda}=\mathbf{W}^\dagger \mathbf{VW}$.
\begin{equation}
    \label{plasmonic2}
    \mathcal{E}_\text{MBD}=\frac{1}{2}\text{Tr}{(\sqrt{\mathbf{V}})}-\frac{3}{2}\sum_{i=1}^N\omega_i
\end{equation}
The evaluation of the trace of a symmetric matrix function such as $\text{Tr}{[\sqrt{\mathbf{V}}]}$ is, in the proposed SL-MBD, based on two main assumptions.\\
First, the stochastic Hutchinson trace estimator (HTE) \cite{hutchinson} is invoked, Eq.\eqref{hutch}, $\mathbf{v}_l$ being one of the $R$ normalized random vectors of dimension $D$ (in our case $D=3N)$, which entries follow a Rademacher distribution, \emph{i.e.} they can assume values of either $1$ or $-1$ with the same probability.
\begin{equation}
    \label{hutch}
  \text{Tr}[\sqrt{\mathbf{V}}]\approx \frac{D}{R}\sum_{l=1}^R\mathbf{v}_l^\dagger\sqrt{\mathbf{V}}\mathbf{v}_l
\end{equation}
\begin{equation}
    \label{dds}
    \begin{split}
    \mathbf{v}_{l}&=\frac{\mathbf{u}_{l}}{\|\mathbf{u}_{l}\|}\\
     u_{l,i}&=
        \begin{cases}
~1,~~~ \textbf{Pr}= 1/2\\
-1,~~ \textbf{Pr}= 1/2
\end{cases}
 \end{split}
\end{equation}
Second, each of the $R$ scalar expectation values in Eq.\eqref{hutch} can be expressed in terms of $\text{Tr}[\sqrt{\boldsymbol{\Lambda}}]$ and the unitary transformation $\mathbf{W}$ as reported in Eq.\eqref{sl1} where we introduced $\boldsymbol{\mu}_{l}=\mathbf{W}^\dagger\mathbf{v}_l$.
\begin{equation}
    \label{sl1}
    \mathbf{v}_l^\dagger \sqrt{\mathbf{V}}\mathbf{v}_l=\mathbf{v}_l^\dagger \mathbf{W}\sqrt{\boldsymbol{\Lambda}}\mathbf{W}^\dagger\mathbf{v}_l=\sum_{i}^D\mu_{l,i}^2\sqrt{\lambda_i}
\end{equation}
The last equality in Eq.\eqref{sl1} corresponds to the Riemann–Stieltjes integral\cite{lanczquadr} defined in Eq.\eqref{sl2} which is approximated via the general $(M+1)$-points quadrature shown in eq.\eqref{sl3}, $\{\tau_k\}$ and $\{\theta_k\}$ representing the unknown weights and nodes respectively.
\begin{equation}
    \label{sl2}
    \begin{split}
    \sum_{i}^D\mu_{l,i}^2\sqrt{\lambda_i}&=\int_a^b \sqrt{t}d\mu(t)\\
         \mu(t)&=
        \begin{cases}
0~~~~~~~~~~~~,~~~ t<a=\lambda_1\\
\sum_{j=1}^{i-1}\mu_j^2~~~,~~~\lambda_{i-1}\leq t < \lambda_i\\
\sum_{j=1}^{D}\mu_j^2~~~,~~~b=\lambda_n<t
\end{cases}
\end{split}
\end{equation}
\begin{equation}
    \label{sl3}
    \mathbf{v}_l^\dagger \sqrt{\mathbf{V}}\mathbf{v}_l=\int_a^b \sqrt{t}d\mu(t)\approx  \sum_{k=1}^{M+1}\tau^{(l)}_{k}\sqrt{\theta_k}
\end{equation}
By inserting Eq.\eqref{sl3} in Eq.\eqref{hutch}, one can identify the complete expression for the stochastic trace estimation, Eq.\eqref{sl4}.
\begin{equation}
    \label{sl4}
    \text{Tr}{[\sqrt{\mathbf{V}}]}\approx \frac{D}{R}\sum_{l=1}^R\sum_{k=1}^{M+1}\tau^{(l)}_{k}\sqrt{\theta_k^{(l)}}
\end{equation}
In the stochastic Lanczos algorithm, the nodes and weights for the quadrature relative to each of the l-th terms in the first summation, are identified as the eigenvalues $\{\widetilde{\lambda}_k^{(l)}\}$ and the first entry (squared) of the eigenvectors $\{[U_{1,k}^{(l)}]^2\}$ of the tridiagonal $\boldsymbol{\Delta}^{(l)}$ matrix which is the representation of the original MBD potential matrix $\mathbf{V}$ in the $M+1$ Krylov subspace $\mathcal{K}_{M+1}=\{\mathbf{y}_1,\mathbf{y}_2,\dots,\mathbf{y}_{M+1}\}$ where the basis vectors are gathered as the $\mathbf{Y}^{(l)}$ matrix's columns.
\begin{equation}
    \label{tridiag1}
    \boldsymbol{\Delta}^{(l)}=\mathbf{Y^\dagger}^{(l)}  \mathbf{VY}^{(l)}
\end{equation}
\begin{equation}
    \label{tridiag2}
    \widetilde{\boldsymbol{\Lambda}}^{(l)}=\mathbf{U}^{(l)\dagger} \boldsymbol{\Delta}^{(l)}\mathbf{U}^{(l)}
\end{equation}
The solution of eq.\eqref{tridiag1} represents the crucial part of the algorithm in terms of efficiency while eq.\eqref{tridiag2}, by virtue of the small matrices involved (Krylov subspace dimension rarely exceeding 15), is inexpensive and it is solved by means of standard libraries. \\
Eq.\eqref{tridiag1} is practically solved as follows: For each of the R terms employed in the HTE, $\mathbf{v}$ (from now on the upperscript $(l)$ is dropped for simplicity) is taken as the first basis vector of the Krylov subspace ($\mathbf{y}_1$) while the remaining basis vectors $\{\mathbf{y}_k \}$ (columns of $\mathbf{Y}$) and the diagonal ($\Delta_{kk}$) and out-of diagonal ($\Delta_{(k-1)k}=\Delta_{k(k-1)}$) elements of $\boldsymbol{\Delta}$ are retrieved recursively as shown in eq.\eqref{recurs} where the asterisk denotes the unnormalized k-th basis vector.

\begin{equation}
\label{recurs}
    \begin{split}
    \mathbf{y}_1&= \mathbf{v}\\
       b_k\mathbf{y}_k&=\mathbf{y^*}_{k}= \mathbf{l}_{k-1}-a_{k-1}\mathbf{y}_{k-1}-b_{k-1}\mathbf{y}_{k-2}\\
       \mathbf{l}_k&=\mathbf{Vy}_k\\
       b_{k}&=\sqrt{\mathbf{y^*}_k^\dagger\mathbf{y^*}_k}=\Delta_{(k-1)k}=\Delta_{k(k-1)}\\
       a_k&=\mathbf{y}_k^\dagger\mathbf{Vy}_k=\mathbf{y}_k^\dagger \mathbf{l}_k=\Delta_{kk}
    \end{split}
\end{equation}
\[
\boldsymbol{\Delta}^{(l)} = \begin{pmatrix}
    \Delta_{11}^{(l)} & \Delta_{12}^{(l)} &  0  &  0   &  0  \\
     \Delta_{21}^{(l)} & \ddots       &  \ddots &    0    & 0\\
     0 &   \ddots    &    \Delta_{kk}^{(l)}     &  \ddots   &  0 \\
     0 &    0   &   \ddots      &     \ddots      &  \Delta_{(M)(M+1)}^{(l)} \\
     0 &   0      &     0      &     \Delta_{(M+1)(M)}^{(l)}      &  \Delta_{(M+1)(M+1)}^{(l)}  \\
     
  \end{pmatrix}
\]
In general the k-th iteration retrieves the $\Delta_{kk}$ diagonal element as well as the contiguous upper/lower $\Delta_{(k-1)k}$ and $\Delta_{k(k-1)}$ ones. In the next section, expressions for $\mathbf{y}$, $a_k$ and $b_k$ in the case of full PBC enforced via PME method will be derived. 
\section{\label{sec:theory}Theory}
\begin{figure}[!htb]{}
   \includegraphics[width=0.82\linewidth]{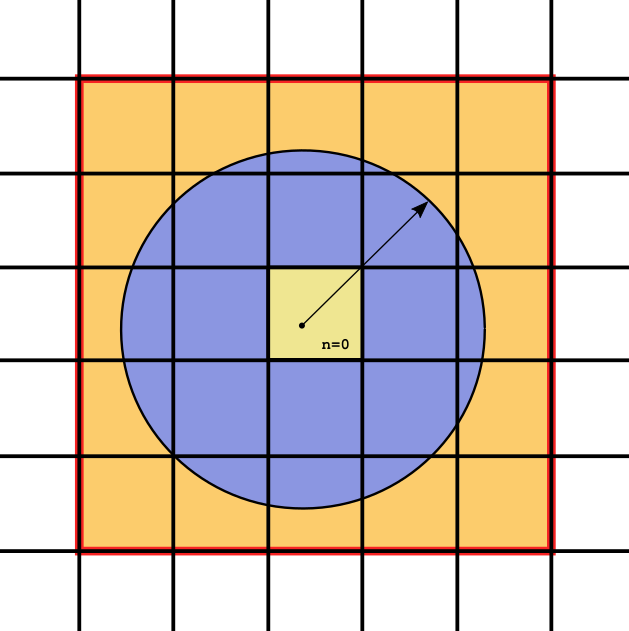}
   \caption{Pictorial representation of the replica method for a 2-D squared box of side $L$ where the chosen cutoff radius is $R_\text{cut}$. Given the ratio $x=R_\text{cut}/L$, a supercell (yellow ochre delimited by red boundary) with vertices identified from all the four possible pairs of integers $(\pm n_\text{max},\pm n_\text{max})$ is built. $n_\text{max}$ represents the smallest integer value that is greater than or equal to $x$. A given particle belonging to the central cell ($\mathbf{n}=\mathbf{0}$) will therefore interact with other particles in supercell placed at a distance smaller than $R_\text{cut}$, i.e. within the blue circle.}
   \label{fig:replica_pic}
\end{figure}

The easiest strategy for including PBC in the MBD model consists in looping over a selected number of cell vectors $\mathbf{n}$, each of which denoting the periodic image of the central simulation cell $U$ defined by its edges $(\mathbf{a}_1,\mathbf{a}_2 ,\mathbf{a}_3 )$ and with volume $V=\mathbf{a}_1 \cdot (\mathbf{a}_2 \times \mathbf{a}_3)$. 
This would result in the modified dipole-dipole interaction matrix $\mathbf{T}^\text{pbc}$ shown in eq.\eqref{replica} where $\mathbf{T'}_{ij}(j\in \mathbf{0})$ represents the $ij$ interaction block belonging to the central simulation cell while $\mathbf{T'}_{ij}(j\in\mathbf{n})$ the interaction between the particle $i$ and the particle $j$ this time belonging to the cell's periodic replica identified with $\mathbf{n}$. In particular, the list of cells (and therefore their associated $\mathbf{n}$ vectors) are chosen according to a cutoff radius as pictorially represented in Fig.\ref{fig:replica_pic} 
\begin{equation}\label{replica}
\begin{split}
\mathbf{T}^\text{pbc}_{ij}&=\mathbf{T'}_{ij}(j\in \mathbf{0})+\sum_{\mathbf{n}\neq\mathbf{0}}\mathbf{T'}_{ij}(j\in\mathbf{n})\\
\mathbf{n}&=n_1\mathbf{a}_1 + n_2\mathbf{a}_2+n_3\mathbf{a}_3~~~~n_1,n_2,n_3\in \mathbb{Z}^3
\end{split}
\end{equation}

The substitution of $\mathbf{T'}_{ij}$ with $\mathbf{T}^\text{pbc}_{ij}$ inside $\mathbf{V}$ (often referred as to the replica method) and the subsequent use of its eigenvalues in eq.\eqref{plasmonic} was discussed in reference.\cite{car_pbc} However, the use of truncated methos based on eq.\eqref{replica} involves the problematics listed and discussed below.\\
First, the summation in eq.\eqref{replica} represents a slowly and conditionally convergent series that characterizes not only dipole-dipole interactions, but also charge-charge, charge-dipole and charge-quadrupole Coulomb interactions kernels.\cite{cond_conv}\\
Consequently, the slow convergence of eq.\eqref{replica} strongly limits the applicability of the SL-MBD algorithm where the efficient ``on-the-fly'' computation of each $\mathbf{V}_{ij}$ block is crucial for the evaluation of the $\mathbf{Vy}_k$ products discussed in connection to eq.\eqref{recurs}.\\
The Ewald summation (ES) method , as well as its more efficient PME variants, was design to improve over eq.\eqref{replica} since the conditionally convergent features of long-range electrostatic interactions of periodic systems are replaced by an absolutely convergent treatment.\\
Let's consider a set of $N$ interacting dipoles belonging to the central simulation cell $U$ and gathered into the $3N$-dimensional array $\mathbf{d}$. The correspondent electric field array $\mathbf{E}=\mathbf{T}^\text{pbc}\mathbf{d}$ arising from the dipoles in both the central simulation cell and all its periodic images is, in the ES method, expressed as the sum of three component, eq.\eqref{ewalddip}.
\begin{equation}
\label{ewalddip}
 \mathbf{E} = \mathbf{T}^\text{pbc}\mathbf{d}\longrightarrow \mathbf{E}^\star=\mathbf{E}^\text{dir}+\mathbf{E}^\text{rec}+\mathbf{E}^\text{self}
\end{equation}
$\mathbf{E}^\text{dir}$ represents the direct space contribution to the Ewald electric field, the $\mathbf{E}^\text{rec}$ is the long-range term computed in Fourier (reciprocal) space while $\mathbf{E}^\text{self}$ represents the so called self-interaction term. The explicit expressions for each of these terms will be given later in the discussion, however, it is important to stress that each of these field components consist of absolutely convergent contributions as the resulting $\mathbf{E}^\star$ field.\\
Our strategy is thus to identify and isolate from the SL-MBD equations, eq.\eqref{recurs}, an electric field-like term that can be then evaluated according to the three absolute convergent contributions in eq.\eqref{ewalddip}, thus allowing us to include PBC in a robust and efficient manner.\\
To do so, we will now start by partitioning $\mathbf{V}$ into its diagonal and out-of-diagonal contributions given below, where $\mathbf{I}_3$ is a (3,3) identity matrix.
\begin{equation}
    \label{vmbd}
    \begin{split}
    \mathbf{V}_{ij}&={\omega}_i{\omega}_j\sqrt{{\alpha}_i(0){\alpha}_j(0)} \mathbf{T'}_{ij} \\
    \mathbf{V}_{ii}&=\mathbf{I}_3 {\omega}^2_i
    \end{split}
\end{equation}
Due to the fact that the diagonal blocks $\mathbf{V}_{ii}$ are themselves diagonal, we introduce the identity in eq.\eqref{ccd}, where $\boldsymbol{\Omega}$ is the diagonal matrix defined below and $\widetilde{\mathbf{V}}$ is the hollow matrix coposed of the off-diagonal entries of $\mathbf{V}$. These quantities will turn useful later in the discussion.
\begin{equation}
    \label{ccd}
    \begin{split}
     \mathbf{V}=&   \boldsymbol{\Omega} + \widetilde{\mathbf{V}}\\
     \boldsymbol{\Omega}=& \bigoplus_i^N \mathbf{V}_{ii}
    \end{split}
\end{equation}
We further introduce the $\mathbf{g}$ vector (of dimension $3N$) defined as the concatenation of $N$ three-dimensional vectors-of-ones ($\mathbf{1}_3$) as shown in Eq.\eqref{gvec}.
\begin{equation}
\label{gvec}
\begin{split}
 \mathbf{g}&=\bigoplus_i^N\omega_i\sqrt{\alpha_i(0)}\mathbf{1}_3
\end{split}
\end{equation}
At this point, we use the newly introduced quantities defined in eq.\eqref{ccd} to rewrite the diagonal $a_k$ term as shown in eq.\eqref{ak}.
\begin{equation}\label{ak}
a_k=\mathbf{y}_k^\dagger\boldsymbol{\Omega}\mathbf{y}_k+\mathbf{y}_k^\dagger\widetilde{\mathbf{V}}\mathbf{y}_k
\end{equation}
One can now easily prove that the second term on the right hand side of eq.\eqref{ak} can be rewritten in terms of the $\mathbf{g}$, eq.\eqref{hadamard}, where $\odot$ denotes the Hadamard product.
\begin{equation}\label{hadamard}
    \mathbf{y}_k^\dagger\widetilde{\mathbf{V}}\mathbf{y}_k=(\mathbf{y}_k\odot\mathbf{g} )^\dagger\mathbf{T'}(\mathbf{g} \odot \mathbf{y}_k)
\end{equation}
By inserting Eq.\eqref{hadamard} into \eqref{ak}, we obtain an expression for $a_k$ which will soon turn crucial for the discussion.
\begin{equation}
    \label{ak2}
    a_k=\mathbf{y}_k^\dagger\boldsymbol{\Omega}\mathbf{y}_k+(\mathbf{y}_k\odot\mathbf{g} )^\dagger\mathbf{T'}(\mathbf{g} \odot \mathbf{y}_k)
\end{equation}
The $3N$-dimensional term $(\mathbf{g} \odot \mathbf{y}_k)$ can be thought as a generalized dipole array $\mathbf{d}_k$ that, via the interaction tensor $\mathbf{T}$ originates the generalized field $\mathbf{E}_k=\mathbf{Td}_k$ that can be then eventually computed according to eq.\eqref{ewalddip}.
\begin{equation}
        \label{ak3}
    a_k=\mathbf{y}_k^\dagger\boldsymbol{\Omega}\mathbf{y}_k+\mathbf{d}_k^\dagger\mathbf{E}^\star_k
\end{equation}
We note in passing that the introduction of this generalized field can be used in different situations as it allows us to couple our system with en external perturbation that, as discussed in references, could arise from implicit solvent contribution.\cite{poierjensen3,poierjensen4}\\ 
At this point we note from eq.\eqref{recurs} (last equality) that $a_k$ is related to $\mathbf{l}_k$ via a differentiation with respect to the basis vector $\mathbf{y}_k$. We can therefore differentiate eq.\eqref{ak2} to finally obtain eq.\eqref{lk2} where the rule for the differentiation of a commuting Hadamard product has been applied, eq.\eqref{haddif}. We note that a similar approach based on differentiation was adopted by Stamm and co-workers in deriving Ewald summation for arbitrary orders of multipoles with particular emphasis on the self term, for which different expressions can be found in literature.\cite{stamm_selfterm} 
\begin{equation}
    \label{haddif}
    \frac{\partial (\mathbf{y}_k\odot\mathbf{g} )}{\partial \mathbf{y}_k}=\frac{\partial\text{Diag}(\mathbf{g})}{\partial \mathbf{y}_k}\mathbf{y}_k+\text{Diag}(\mathbf{g})\frac{\partial\mathbf{y}_k}{\partial \mathbf{y}_k}=\text{Diag}(\mathbf{g})
\end{equation}
\begin{equation}
    \label{lk2}
    \mathbf{l}_k=\frac{1}{2}\frac{\partial a_k}{\partial\mathbf{y}_k}=\boldsymbol{\Omega}\mathbf{y}_k+\text{Diag}(\mathbf{g})\mathbf{T'} (\mathbf{g} \odot \mathbf{y}_k)
\end{equation}
Once again we use the definition of the generalized dipole and field to finally obtain eq.\eqref{lk3}.
\begin{equation}
    \label{lk3}
    \begin{split}
     \mathbf{l}_k=&  \boldsymbol{\Omega}\mathbf{y}_k+ \text{Diag}(\mathbf{g})\mathbf{T'd}_k\\
     =&  \boldsymbol{\Omega}\mathbf{y}_k+ \text{Diag}(\mathbf{g})\mathbf{E}^\star_k
    \end{split}
\end{equation}
Eq.\eqref{recurs} can therefore be rewritten in terms of the generalized electric field $\mathbf{E}_k^\star$ through the above derived quantities, eq.\eqref{recurs2}.
\begin{equation}
\label{recurs2}
    \begin{split}
    \mathbf{y}_1&= \mathbf{v}\\
       b_k\mathbf{y}_k&=\mathbf{y^*}_{k}= \mathbf{l}_{k-1}-a_{k-1}\mathbf{y}_{k-1}-b_{k-1}\mathbf{y}_{k-2}\\
       \mathbf{l}_k&=\boldsymbol{\Omega}\mathbf{y}_k+ \text{Diag}(\mathbf{g})\mathbf{E}^\star_k\\
       b_{k}&=\sqrt{\mathbf{y^*}_k^\dagger\mathbf{y^*}_k}=\Delta_{(k-1)k}=\Delta_{k(k-1)}\\
       a_k&=\mathbf{y}_k^\dagger\boldsymbol{\Omega}\mathbf{y}_k+\mathbf{d}_k^\dagger\mathbf{E}^\star_k=\Delta_{kk}
    \end{split}
\end{equation}
$\mathbf{E}^\star_k$ can be evaluated by ES and the explicit expressions for $\mathbf{E}^\text{dir}$, $\mathbf{E}^\text{self}$ and $\mathbf{E}^\text{rec}$  are shown below, however , for a broader discussion and derivation we refer to the following references.\cite{cond_conv,ewald1,revcomp24}\\
 Starting from the direct component, we identify the three dimensional electric field $\Vec{\mathbf{E}}_{i,k}^\text{dir}$ at the atomic position $\mathbf{R}_i$ arising from the generalized dipole array $\mathbf{d}_k$, where its three-dimensional contribution related to the $j$-th atom is denoted $\vec{\mathbf{d}}_{j,k}$, as shown in eq.\eqref{fld_ewald}.
\begin{equation}
\label{fld_ewald}
\begin{split}
L_{j,k}&=\vec{\mathbf{d}}_{j,k}\nabla_j\\
\Vec{\mathbf{E}}_{i,k}^\text{dir}&= -\sum_{\mathbf{n}}\sum_{j=1}^{N^*} L_{j,k}  \frac{\partial}{\partial \mathbf{R}_i}\biggl(\frac{\text{erfc}(\tau \mid \mathbf{R}_j-\mathbf{R}_i + \mathbf{n} \mid)}{\mid \mathbf{R}_j-\mathbf{R}_i + \mathbf{n} \mid}\\
& ~~~~~~+\sum_{j=1}^{N^*}L_{j,k}(1-s_{ij})\mathbf{T}_{ij}\Vec{\mathbf{d}}_{j,k}\biggr)
\end{split}
\end{equation}
 In the above, $\tau$ represents a real parameter governing the balance between the direct and reciprocal contributions. For a cubic cell of side $h$, it is typically taken to be $5/h$.\cite{allen} $\tau$ is commonly chosen so that the direct term convergence is fast as the reciprocal contribution can be efficiently computed via FFT.  This makes the summation over $\mathbf{n}$ fastly converging, and only particles belonging to neighboring periodic images are therefore usually considered. $\mathbf{E}^\text{dir}$ is practically computed by means of neighbor lists based on the choice of $\tau$ determining the suitable cutoff and this ensures an efficient and linear-scaling evaluation.\\
 The self term $\Vec{\mathbf{E}}_{i,k}^\text{self}$ consists in the single term shown in eq.\eqref{fld_ewald1} which evaluation involves a negligible computational effort.\\
\begin{equation}
\label{fld_ewald1}
\Vec{\mathbf{E}}_{i,k}^\text{self}= \frac{2\tau^3}{3\sqrt{\pi}}\Vec{\mathbf{d}}_{i,k}
\end{equation}
From a computational point of view, with standard $\tau$ parameters, the most expensive and thus crucial term to evaluate is represented by the $\Vec{\mathbf{E}}_{i,k}^\text{rec}$ contribution. In order to discuss its explicit expression, we introduce the reciprocal conjugate vectors $(\mathbf{a}^*_1,\mathbf{a}^*_2 ,\mathbf{a}^*_3 )$ which are related to their dual set by $\mathbf{a}^*_\alpha \cdot \mathbf{a}_\beta=\delta_{\alpha \beta}$, with $\alpha,\beta = \{1,2,3\}$ and $\delta_{\alpha \beta}$ being the Kronecker delta. In analogy to what done for $\mathbf{n}$, we define $\mathbf{m}$.
\begin{equation}
\label{dir_rec_intvec}
\mathbf{m}=m_1\mathbf{a}^*_1 + m_2\mathbf{a}^*_2+m_3\mathbf{a}^*_3~~~m_1,m_2,m_3 \in \mathbb{Z}^3
\end{equation}
We further introduce the structure factor $S(\mathbf{m})$, defined in eq \eqref{structurefactor} for a given $\mathbf{m}$ is defined in .
\begin{equation}
\label{structurefactor}
S(\mathbf{m})=\sum_{j=1}^N \vec{\mathbf{d}}_{j,k}\cdot\mathbf{m} \exp{(2i\pi \mathbf{m}\cdot \mathbf{R}_j)}
\end{equation}
In the Ewald summation method the reciprocal component of the field is finally given in eq.\eqref{fld_ewald2}
\begin{equation}
\label{fld_ewald2}
\Vec{\mathbf{E}}_{i,k}^\text{rec}=-\frac{1}{\pi V} \sum_{\mathbf{m}\neq \mathbf{0}} \frac{\partial}{\partial \mathbf{R}_i}\biggl(\frac{\exp{(-\pi^2 \mathbf{m}^2/\tau^2})}{\mathbf{m}^2} S(\mathbf{m})\exp{(-2i\pi\mathbf{m}\cdot \mathbf{R}_i)}\biggr)
\end{equation}
The optimal choice of $\tau$ makes the evaluation of eq.\eqref{fld_ewald2} (and therefore of the whole ES method) $\mathcal{O}(N^{3/2})$ scaling, however, the PME method sensibly improves the scaling by approximating the complex exponentials via interpolation. In the Smooth PME method (SPME) in particular, the complex exponentials are first rewritten in terms of the scaled fractional coordinates $u_{\alpha j}$, eq \ref{scaledfractional}, and then interpolated by a $p$-degree B-spline function  $\theta_p(u_{\alpha j}-n_\alpha )$ on a grid of size $K_1\times K_2 \times K_3$ and the final contribution due to the reciprocal space is given in eq.\eqref{spme}
\begin{equation}
\label{scaledfractional}
\begin{split}
&u_{\alpha j}=K_\alpha \mathbf{a}_\alpha^*\cdot \mathbf{R}_j~~~~~~~~~~~~~~~\alpha = \{1,2,3\}~~,~~K_\alpha \in \mathbb{N}^+ \\
&\exp{(2i\mathbf{m}\cdot \mathbf{R}_j)}=\prod _{\alpha=1}^3\exp{\biggl(2i\pi m_\alpha\frac{u_{\alpha_j}}{K_\alpha}}\biggr)
\end{split}    
\end{equation}
The $\Vec{\mathbf{E}}_{i,k}^\text{rec}$ is finally given by eq.\eqref{spme}.
\begin{equation}
\label{spme}
\Vec{\mathbf{E}}_{i,k}^\text{rec}\approx -\frac{\partial}{\partial \mathbf{R}_i}\sum_{\mathbf{n}}\prod_{\alpha=1} ^3 \theta_p (u_{\alpha,i}-n_\alpha)(G^R * D^R)(\mathbf{n})
\end{equation}
The $(G^R * Q^R)$ term is the convolution between the pair potential $G^R$ discussed by Sagui \emph{et al.}  and the real space dipole array $D^R$ defined below.\cite{sagui}  The use of fast Fourier transtorms in the evaluation of \eqref{spme} ensures an overall $\mathcal{O}(N\log(N))$ scaling. 
\begin{equation}
    \label{dipolarr}
    \begin{split}
    D^R_k(k_1,k_2,k_3)=\sum_{\mathbf{n}} &\sum_jL_{j,k} \theta_p(u_{1,j-k_1-K_1n_1})\theta_p(u_{2,j-k_2-K_2n_2})\\ &\theta_p(u_{3,j-k_3-K_3n_3})
    \end{split}
\end{equation}

    \begin{algorithm}[H]
	\caption{Schematic general representation of the SPME-based stochastic Lanczos algorithm.} 
	\label{alg:SL}
	\begin{algorithmic}[1]
     
	    \State SPME Grid allocation $(K_1,K_2,K_3)$ and initialization
     \State Neighbor list (direct space) generation
     
		\For {$({l} =1, R)$}
        \State Generate $\mathbf{v}_l$ from a Rademacher distribution
        \State  $\mathbf{y}^{(l)}_1=\mathbf{v}_l$
        \State call SPME-Lanczos ($k=1$): $\Delta^{(l)}_{11}$
		 \For {$({k} =2, M+1)$}
		 \State call SPME-Lanczos (general $k$): $ \mathbf{y}^{(l)}_k~,~a^{(l)}_{k}~,~b^{(l)}_{k}$
		 \EndFor
		 \State Eigendecomposition of $\boldsymbol{\Delta}^{(l)}:  \mathbf{U}^{\dagger (l)} \boldsymbol{\Delta}^{(l)}\mathbf{U}^{(l)}=\widetilde{\boldsymbol{\Lambda}}^{(l)}$
          \For {$({k} =1, M+1)$}
          \State $E_\text{sum}=\text{E}_\text{sum}+[U^{(l)}_{1,k}]^2\sqrt{\widetilde{\lambda}^{(l)}_k}  $
          \EndFor
          \EndFor
          \State Calculate the average over samples: $\text{Tr}{[\sqrt{\mathbf{V}}]}\approx \frac{3N}{R}E_{sum}$
	\end{algorithmic} 
\end{algorithm}
   \begin{algorithm}[H]
	\caption{Schematic representation of the k-th iteration SPME-Lanczos routine} 
	\label{alg:SLd}
	\begin{algorithmic}[1]
     
	    \State \textbf{Input quantities }$\longleftarrow$ $\mathbf{y}_{k-1}$,$\mathbf{y}_{k-2}$,$\mathbf{l}_{k-1}$,$a_{k-1}$,$b_{k-1}$
       \State build $\mathbf{y}^*_k$ from input quantities
        \State get $b_k$ from the normalization of $\mathbf{y}^*_k$ 
        \State build the generalized dipole vector $\mathbf{d}_k$ from $\mathbf{y}_k$ 
        \State distribute $\mathbf{d}_k$ in the $(K_1,K_2,K_3)$ grid (necessary for $\mathbf{E}_k^\text{rec}$)
        \State compute the generalized field $\mathbf{E}_k^\star=\mathbf{E}_k^\text{dir}+\mathbf{E}_k^\text{self}+\mathbf{E}_k^\text{rec}$
        \State compute $\mathbf{l}_k$ from $\mathbf{E}_k^\star$
		\State compute $a_k$
        \State \textbf{Output quantities } $\longrightarrow$ $\mathbf{y}_{k}$, $b_k$, $\mathbf{l}_k$,$a_k$
	\end{algorithmic} 
\end{algorithm}
The above algorithm was implemented in the Tinker-HP molecular dynamics package\cite{tinkerhp,adjoua2021tinker} and will, in the following section, be numerically analyzed. The replica method (eq.\eqref{replica}) has also been implemented and coupled to the SL-MBD method as this will allow us to perform a direct comparison for a few test cases with the newly proposed SPME version which numerical results will always refer to a fixed Ewald's $\tau$ parameter ($\tau=0.544590$) corresponding to a real space cutoff of 7 Ångstrom.
\section{\label{sec:results}Numerical results}
\begin{figure}[!htb]{}
   \includegraphics[width=0.950\linewidth]{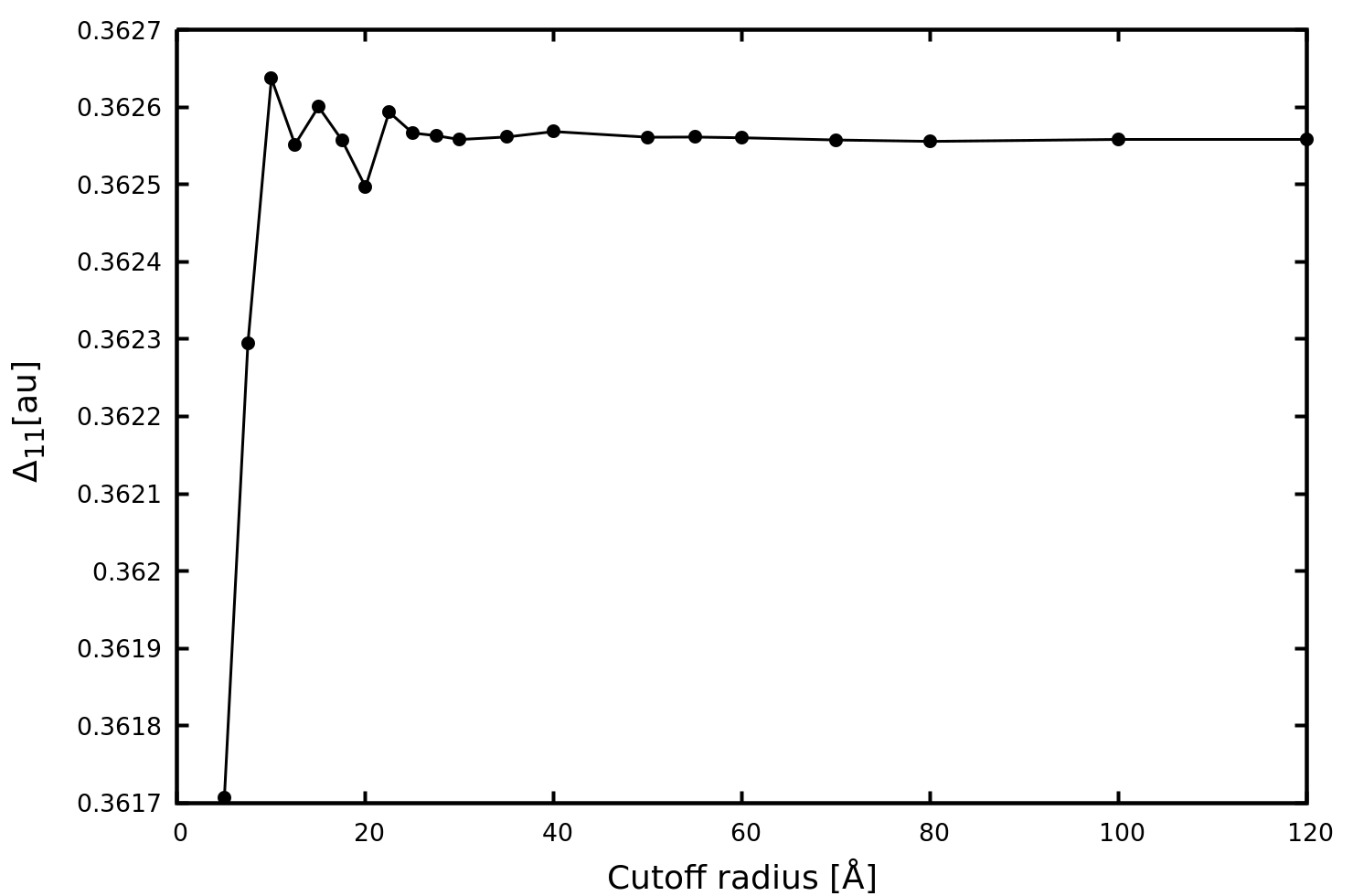}
   \caption{First diagonal element of $\boldsymbol{\Delta}$ computed via the replica method as a function of the cutoff radius for the cubic water box taken as test system.}
   \label{fig:h2o_radius}
\end{figure}

\begin{figure}[!htb]{}
   \includegraphics[width=0.950\linewidth]{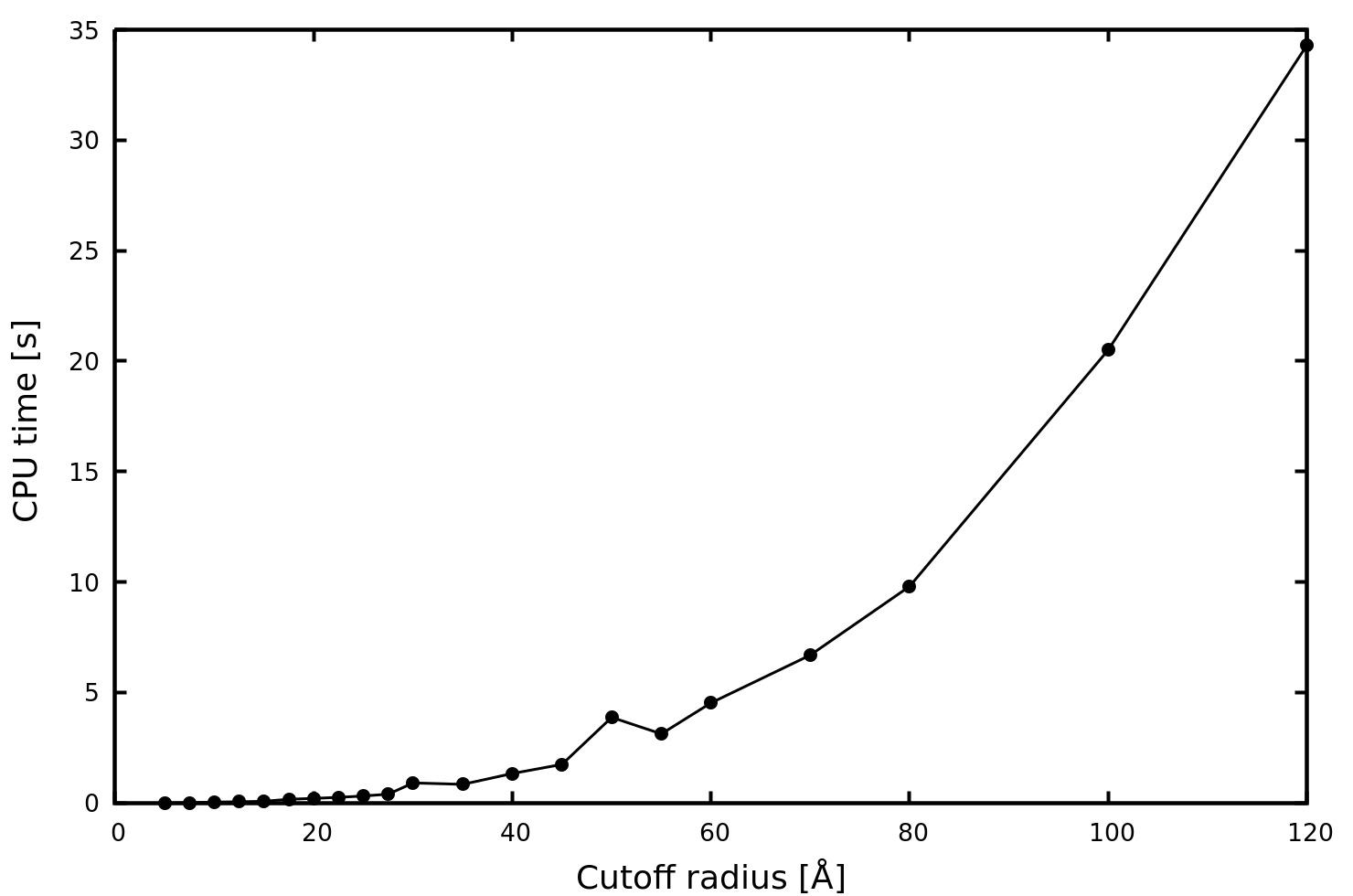}
   \caption{CPU time as a function of the cutoff radius relative to Fig.\ref{fig:h2o_radius}, i.e. waterbox.}
   \label{fig:h2o_replicacpu}
\end{figure}

\begin{figure}[!htb]{}
   \includegraphics[width=0.950\linewidth]{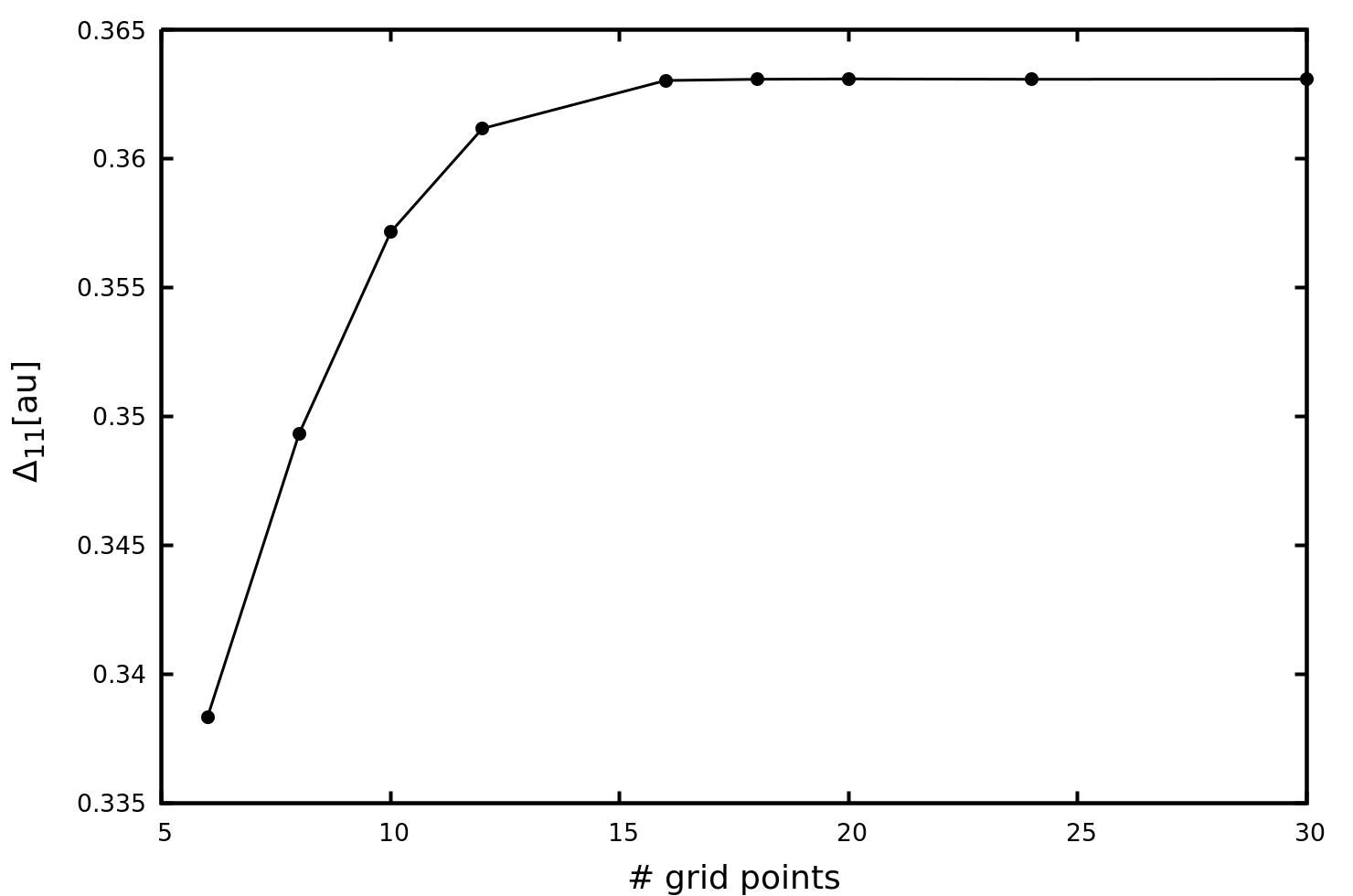}
   \caption{First diagonal element of $\boldsymbol{\Delta}$ computed via the SPME-Lanczos method as a function of the grid points for the waterbox considered(only $K_1$ is reported as the box is cubic). The initial Krylov subspace basis vector $\mathbf{y}_1$ is the same as for the Fig.\ref{fig:h2o_spme}}
   \label{fig:h2o_spme}
\end{figure}

\begin{figure}[!htb]{}
   \includegraphics[width=0.950\linewidth]{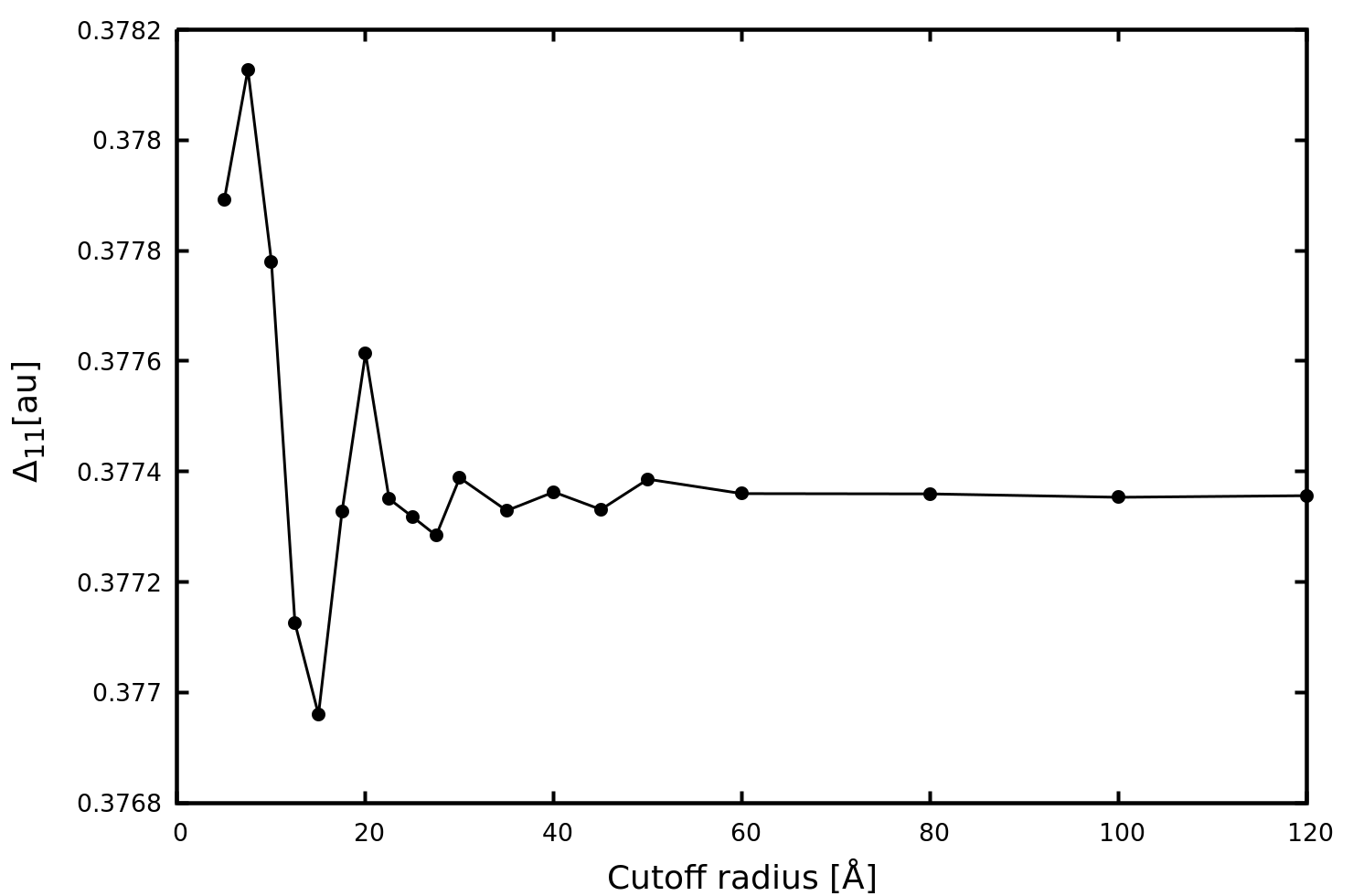}
   \caption{First diagonal element of $\boldsymbol{\Delta}$ as a function of the cutofff radius, computed via the replica method for a cubic box of diamond (14.2 Ångstrom) }
   \label{fig:diamond_replica}
   \end{figure}

\begin{figure}[!htb]{}
   \includegraphics[width=0.950\linewidth]{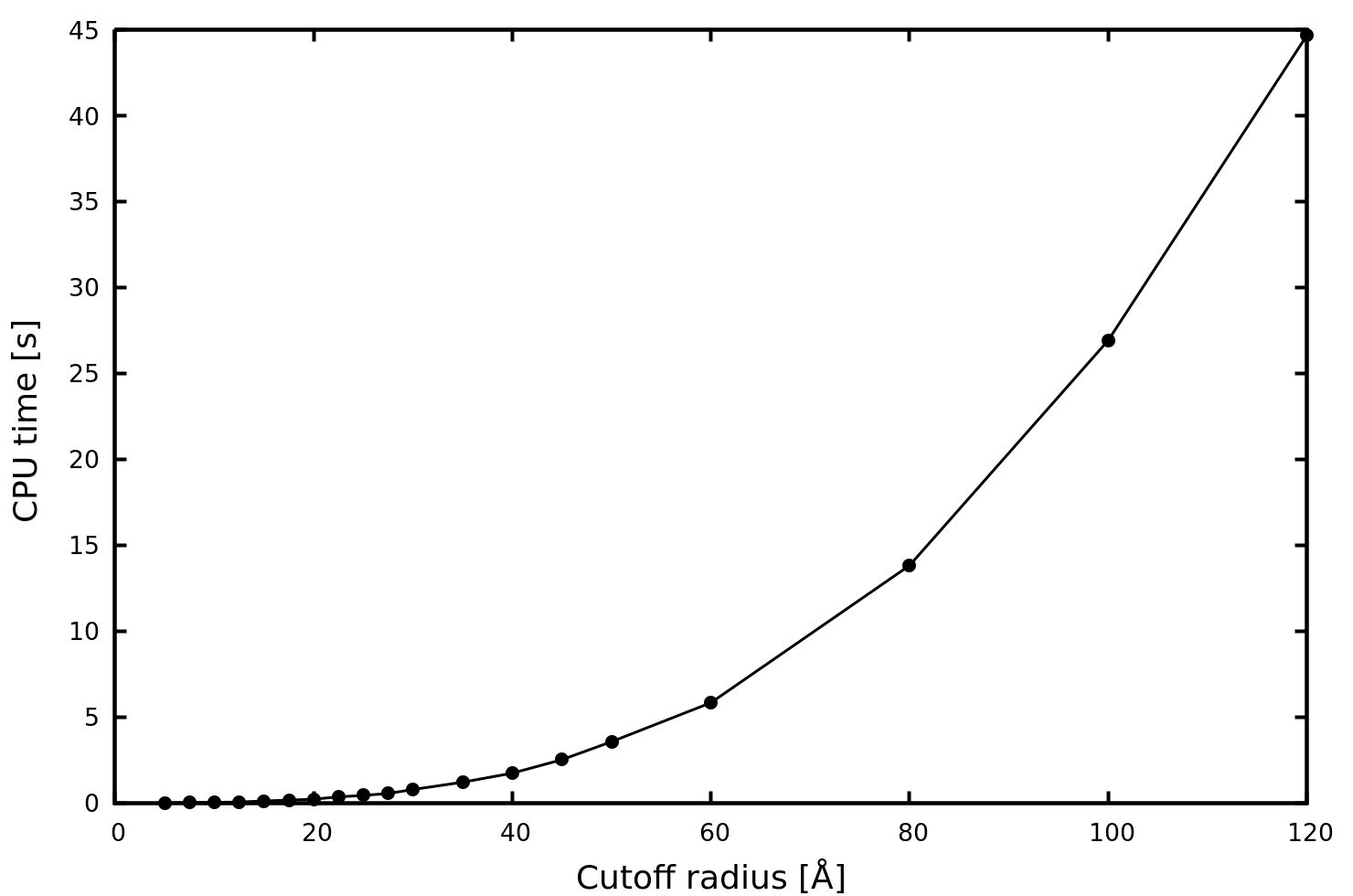}
   \caption{CPU time as a function of the cutoff radius for the replica method for diamond in a cubic box of side 14.2 Ångstrom.}
   \label{fig:diamond_replica_cpu}
   \end{figure}

\begin{figure}[!htb]{}
   \includegraphics[width=0.950\linewidth]{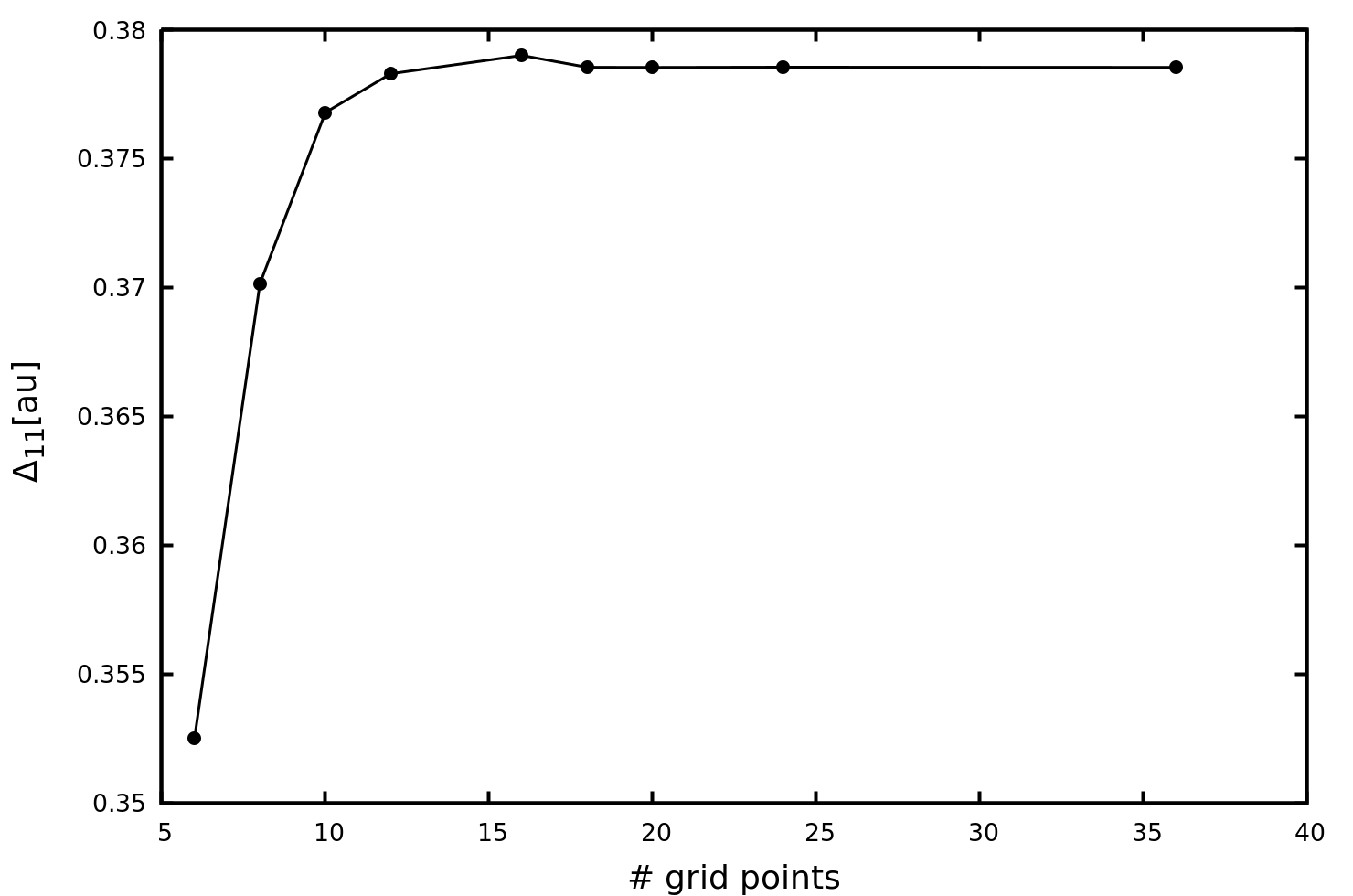}
   \caption{First diagonal element of $\boldsymbol{\Delta}$ computed via the SPME-Lanczos method as a function of the grid points (only $K_1$ is reported as the box is cubic) for diamond in a box of side 14.2 Ångstrom.}
   \label{fig:diamond_spme}
   \end{figure}

\begin{figure}[!htb]{}
   \includegraphics[width=0.950\linewidth]{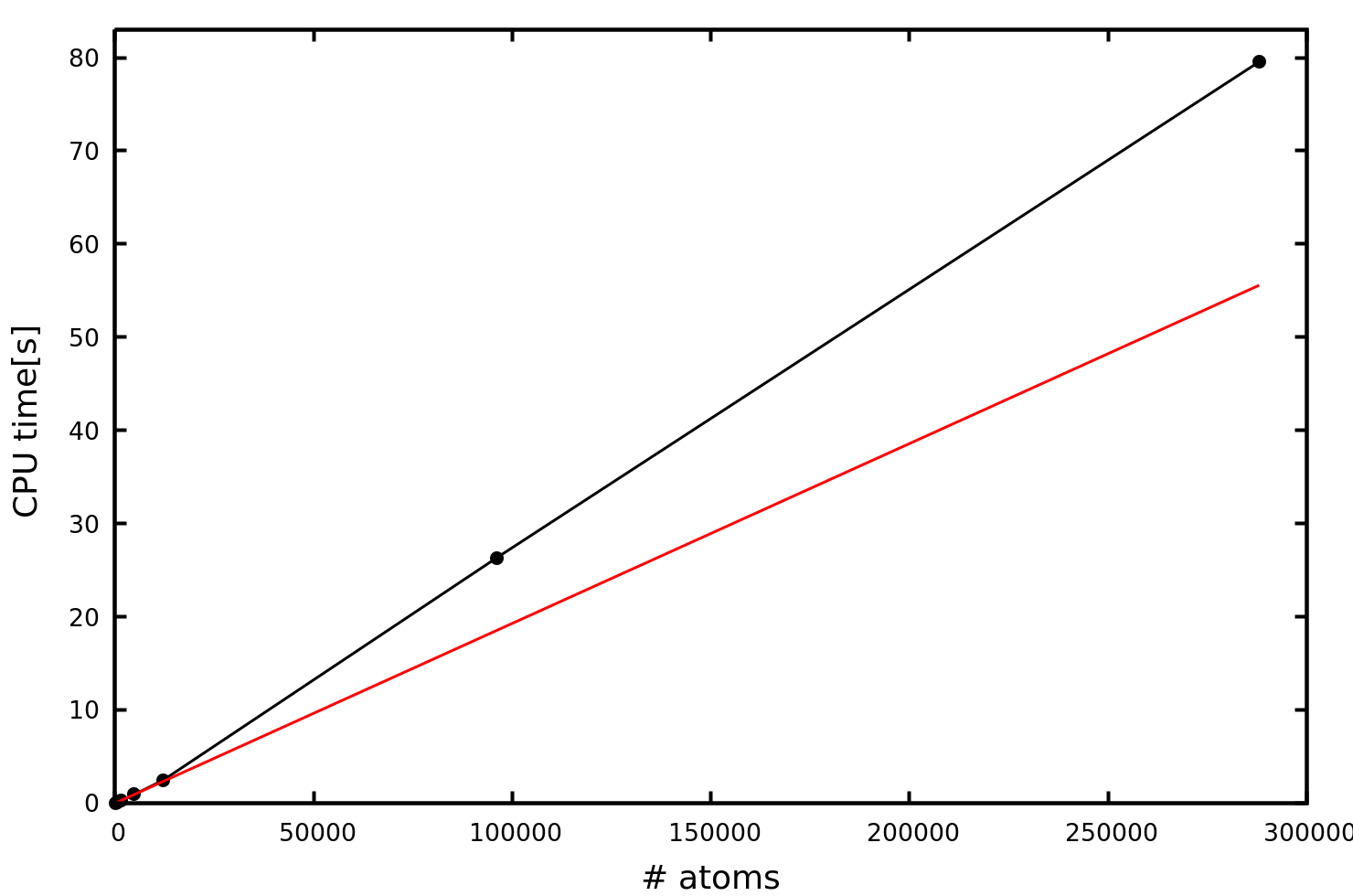}
   \caption{CPU time in seconds as a function of the number of atoms of increasingly larger water boxes (black line). The time refers to the 10 iterations required to build $\boldsymbol{\Delta}$ for a given random sample within a standard $\mathcal{K}_{10}$ Krylov subspace with the implemented SPME-SL algorithm. The red line represents the ideal linear scaling.}
   \label{fig:scaling}
   \end{figure}

We start by considering results related to the simple replica method based on Eq.\eqref{replica}. In particular, for all the results we choose as a measure the first diagonal element of the $\boldsymbol{\Delta}$ matrix calculated from the same fixed initial vector $\mathbf{y}_1=\mathbf{v}$, chosen as usual from a Rademacher distribution. This choice will allow us to eliminate the stochastic noise from the computed $\Delta_{11}$ values that otherwise would make harder the interpretation and comparison of the effects arising from long-range interactions introduced via both the replica and SPME methods. The first system analyzed is a small cubic box of dimension $18.64$ Ångstrom containing 216 water molecules in the liquid phase. Figure \ref{fig:h2o_radius} shows the evolution of $\Delta_{11}$  as a function of the cutoff radius $R_{cut}$ that is used to determine the replicas identified by a set of $\{\mathbf{n}\}$ vectors to be included in eq.\eqref{replica}. 
Even for a not highly symmetric system such as bulk water, the convergence is reached for a cutoff radius of nearly 30 Ångstrom thus confirming the slow (and conditional) convergence rate that characterizes the replica method. The large cutoff radius required by the replica method, because of  its consequent quadratic scaling, has a direct impact on the computational time as shown in Fig.\ref{fig:h2o_replicacpu}.
In particular, for a 30 Ångstrom cutoff the CPU-time required for the computation of the diagonal element chosen as observable reaches 1 second. \\
The situation if quite different if the SPME-based algorithm is employed since in this case the overall convergence is determined by the number of grid points to be used in the solution of the reciprocal field contribution ($K_1,K_2,K_3$ in eq.\eqref{scaledfractional}) that also represents the computationally most expensive part of the algorithm as the direct summation part is computed very efficiently in a linear-scaling fashion. Fig.(\ref{fig:h2o_spme}) shows the convergence of our target quantity $\Delta_{11}$ as a function of the number of grid points for the box of water undertaken as test system. We stress that, given the choice that we made to fix Ewald's $\tau$ parameter, the only quantity governing the convergence is thus the grid size. 
We first note that the convergence has a monotonic behavior as a smaller grid size does not involve a physical truncation of the space and thus of the interactions as for the replica case that in fact shows an oscillatory behavior.\\
It is now interesting to compare the computational cost required by the SPME-based approach to that of the replica method. In particular for a 18 point sized grid for which convergence is observed, the CPU time is $10^{-2}$, therefore a factor 100 faster than the cumbersome replica method. \\
The slow convergence rate observed for the replica method is further exacerbated when highly symmetrical systems are taken under consideration. Fig (\ref{fig:diamond_replica}) shows the evolution of $\Delta_{11}$ as a function of the cutoff radius, this time for a 14.2 Ångstrom sided cubic box of diamond.
In this case the cutoff radius reaches the extremely large value of 60 Ångstrom before the convergence is reached, with a huge impact on outcoming computational cost as shown in Fig\ref{fig:diamond_replica_cpu}.
The system dependency affecting the choice for a proper cutoff radius observed for the replica method is not suffered by the SPME-Lanczos method as it can be seen in Fig.\ref{fig:diamond_spme} showing the $\Delta_{11}$ convergence as a function of the number of grid points.
Even in this case, convergence is observed starting from circa 20 points, similarly to that observed for water as both boxes have quite similar size. In fact, convergence is ensured when a certain density of grid points is provided, independently of the system. In general a density of 1.2 points/Ångstrom (for each of the three box dimensions) is enough to ensure convergence, and this is the default value chosen in our implementation.\\
For highly periodic systems for which the replica method is particularly slow to converge, the computational gain provided by the SPME alternative becomes even more marked. For a 18 points grid, the $\Delta_{11}$ computation  via the SPME-Lanczos is nearly 350 times faster than its replica counterpart. \\
Although our analysis focused, for the sake of clarity, on the $\Delta_{11}$, the same results hold for the convergence of out of diagonal terms $\Delta_{(k-1)(k)}$. Moreover, we note that the solution of the SPME-Lanczos equations, Eq.\eqref{recurs}, does not spoil the orthogonality of the Krylov subspace basis vectors $\mathbf{y}^\dagger_j\mathbf{y}_k=\delta_{jk}$ as the set of vectors remain orthogonal by construction as in the original algorithm. Furthermore, we stress that for an accurate resolution of eq.\eqref{sl2}, the number of quadrature points i.e. the dimension $(M+1)$ of the Krylov subspace $\mathcal{K}_{M+1}$, can be set to 15, regardless of the system size. This implies that the SPME-Lanczos algorithm does not suffer from the numerical instability (loss of orthogonality among basis vectors) of standard Lanczos algorithm\cite{lanczos1950}, typically encountered in applications where very large Krylov subspaces and thus basis vectors are required.\cite{lancz_orth} \\
Being the construction of the tridiagonal matrix $\boldsymbol{\Delta}$ the bottleneck step of the overall algorithm, it is of interest to probe its scaling as a function of the system size as shown in Fig.\ref{fig:scaling} for an increasingly large box of liquid water.\\
The plot shows that the SPME-SL algorithm deviates from linearity for larger system sizes and this is explained by virtue of the $N\log(N)$ scaling of the SPME method employed to compute the generalized field vectors which are key contributions in the construction of the $\boldsymbol{\Delta}$ matrix. The deviation from linearity is, however, rather contained even for the largest system considered composed of approximately 100000 water molecules that is completely out of reach for the standard replica method discussed earlier. \\
For one single core, the overall time necessary to compute the final energy) is equivalent to the time required to build $\boldsymbol{\Delta}$ (Fig.\ref{fig:scaling}) multiplied by the number of random samples $R$ involved in Hutchinson's trace estimator. For large systems in the order of $10^4$ atoms or above, $R$ can be taken to be around 300 with a resulting low relative standard deviation (0.5\%) as analyzed in depth in reference.\cite{mbd_stoch} \\
However, the SPME-SL algorithm's strength is found in its embarrassingly parallel nature since the random samples can be divided
among the available processes while a simple reduction is required before the final trace evaluation (eq.\eqref{sl4}). Since the parallelization scheme is essentially the same as the one discussed in the original SL-MBD algorithm, we refer to a previous work\cite{mbd_stoch} for an in depth analysis of the scalability with respect to the number of processes as well as a detailed discussion of the parallelization strategy.\\

\section{\label{sec:conclusions}Conclusions}
We have derived, implemented an discussed the SPME-SL algorithm where the stochastic Lanczos trace estimation scheme is coupled to the state of the art Smooth Particle Mesh Ewald method. This was made possible by introducing the generalized field term contribution from the Lanczos iterative equations. Our combined approach allows for an embarrassingly parallel  computation of many-body dispersion energies with the full inclusion of long-range interactions arising from all periodic images of the central simulations cell. \\
The proposed algorithm undoubtedly outperforms truncation-based approaches such as the replica method that is affected by slow and conditionally convergence as well as by the employed quadratic-scaling double loops  making the computation highly inefficient for large systems.\\
The parallelism features of the SPME-SL algorithm together with the $N\log(N)$ scaling with the system size allows for a fast Many-Body dispersion treatment of very large periodic systems composed of hundreds of thousands atoms and more. This work represents the natural extension to long-range PBC of our recent stochastic Lanczos MBD algorithm\cite{mbd_stoch} and it focuses uniquely on the energy evaluation. Our focus will now be dedicated to the extension of these achievements to the energy nuclear gradients towards large scale condensed phase molecular dynamics simulations including many-body dispersion effects. Furthermore, alternative handling of PBC within the recently introduced ANKH linear scaling framework will pursued\cite{chollet2023ankh}.

\begin{acknowledgments}
This work has been funded by the European Research Council
(ERC) under the European Union’s Horizon 2020 research
and innovation program (grant No 810367), project EMC2
(JPP). Computations have been performed at GENCI (IDRIS,
Orsay, France and TGCC, Bruy\`eres le Châtel) on grant no
A0070707671.
\end{acknowledgments}

\section*{Conflict of interest}
The authors have no conflicts to disclose.

\bibliography{aipsamp}

\end{document}